\documentclass[namedreferences]{SolarPhysics}
\usepackage[optionalrh,solaenum]{spr-sola-addons} 
\usepackage{graphicx}        
\usepackage{solaheader}
\usepackage{color}           
\usepackage{url}             

\newcommand{\mbf}{\mathbf}

\newcommand{\beq}{\begin{equation}} 
\newcommand{\eeq}{\end{equation}} 
\newcommand{\bfig}{\begin{figure}} 
\newcommand{\efig}{\end{figure}} 
\newcommand{\bit}{\begin{itemize}} 
\newcommand{\eit}{\end{itemize}} 
\newcommand{\ben}{\begin{enumerate}} 
\newcommand{\een}{\end{enumerate}} 
 
\begin{document} 
 
\begin{article} 
 
\begin{opening} 
 
\title{Are Solar Active Regions with Major Flares More Fractal, 
  Multifractal, or Turbulent than Others? }  
 
\author{Manolis K. Georgoulis} 
\runningauthor{Georgoulis} 
\runningtitle{Are Flaring Active Regions More Complex than Others?} 
 
\institute{Research Center for Astronomy and Applied Mathematics (RCAAM),  
           Academy of Athens, 4 Soranou Efesiou Street, Athens, Greece, GR-11527\\ 
                     email: \url{manolis.georgoulis@academyofathens.gr} \\} 
 
\begin{abstract} 
Multiple recent investigations of solar magnetic field measurements have
raised claims that the scale-free (fractal) or multiscale (multifractal)
parameters inferred from the studied magnetograms may help assess the eruptive
potential of solar active regions, or may even help predict major flaring
activity stemming from these regions. We investigate 
these claims here, by testing three widely used scale-free 
and multiscale parameters, namely, the fractal dimension, the multifractal 
structure function and its inertial-range exponent, and the turbulent 
power spectrum and its power-law index, on a comprehensive data set of 
370 timeseries of active-region magnetograms ($17,733$ magnetograms in total) 
observed by SOHO's {\it Michelson Doppler Imager} (MDI) 
over the entire Solar Cycle 23.  
We find that both flaring and non-flaring active regions exhibit significant
fractality, multifractality, and non-Kolmogorov turbulence but none of the
three tested parameters manages to distinguish active regions with major flares
from flare-quiet ones. We also find that the multiscale parameters, but not 
the scale-free fractal dimension, depend sensitively on the spatial
resolution and perhaps the observational characteristics of the studied
magnetograms. Extending previous works, we  
attribute the flare-forecasting inability of 
fractal and multifractal parameters to {\it i)} a widespread 
multiscale complexity caused by a possible underlying 
self-organization in turbulent solar magnetic structures,  
flaring and non-flaring alike, and {\it ii)} a lack of correlation between 
the fractal properties of the photosphere and overlying layers, where solar 
eruptions occur. However useful for understanding solar magnetism, therefore,   
scale-free and multiscale measures may not be optimal tools for
active-region characterization in terms of eruptive ability or, ultimately,
for major solar-flare prediction. 
\end{abstract} 
\end{opening} 
\section{Introduction} 
\label{S-Intro}  
The ever-increasing remote-sensing capabilities of modern solar 
magnetographs have led to the undisputed conclusion that solar (active 
region in particular) magnetic fields exhibit an intrinsic 
complexity. ``Complexity'' is a term commonly used to 
describe an array of properties with one underlying
characteristic: a scale-invariant, self-similar (fractal) or multiscale 
(multifractal) behavior. The measured photospheric magnetic fields in 
active regions are indeed {\it multifractal} ({\it e.g.} 
\opencite{Lawrence_etal93}; \opencite{Abramenko_05}),  
that is, consisting of a number of fractal 
subsets. As such, they are also {\it fractal},  
with a fractal dimension equal to the maximum fractal dimension of the 
ensemble of fractal subsets.  
 
Fractality is a mathematical property  
but with important physical implications.   
Scale-free or multiscale manifestations are 
thought to stem from an underlying self-organized, or 
self-organized critical (SOC), evolution in active 
regions. Self-organization refers to the internal,  
intrinsic reduction of the various parameters (also called degrees of freedom)  
of a nonlinear dynamical system, such as a solar active region,   
into a small number of {\it important}  
parameters that govern the system's evolution and, 
perhaps, its dynamical response \cite{Nicolis_Prigogine89}. 
Assumptions on the nature of just these important 
parameters can lead to models of active-region emergence and 
evolution encapsulated in simplified {\it cellular automata} models 
\cite{Wentzel_Seiden92,Seiden_Wentzel96,Vlahos_etal02,Fragos_etal04}.  
Self-organized criticality, on the other hand, implies that the 
self-organized system evolves through a sequence of metastable states 
into a state of {\it marginal} stability with respect to a critical 
threshold. Local  
excess of the threshold gives rise to spontaneous, intermittent  
instabilities lacking a characteristic size  
(\opencite{Bak_etal87}; \opencite{Bak_96}).  
 
The intrinsic self-organization in solar active regions may 
be attributed to the {\it turbulence} dominating  
the emergence and evolution of solar magnetic fields.  
Tangled, fibril magnetic fields rising from the convection 
zone can be explained via Kolmogorov's theory of fluid 
turbulence ({\it e.g.}, \opencite{Brandenburg_etal90}; \opencite{Longcope_etal98};  
\opencite{Cattaneo_etal03}, and others). Turbulence in the generation and 
ascension of solar magnetic fields leads to turbulent 
photospheric flows ({\it e.g.}, \opencite{Hurlburt_etal95}). Thus, the turbulent 
photosphere is viewed as a driver that gradually but constantly 
perturbs an emerged magnetic-flux system, such as an active  
region, dictating self-organization in it and possibly forcing it toward  
a marginally stable, SOC state ({\it e.g.}, \opencite{Vlahos_Georgoulis04}).  
Turbulent action does not cease in the photosphere, but it extends into the  
solar corona. However, coronal low-$\beta$ turbulence 
may not be the Kolmogorov fluid turbulence 
applying to the high-$\beta$ plasma of the convection 
zone and the photosphere. Instead, it might be an intermittent  
magnetohydrodynamic (MHD) turbulence \cite{Kraichnan_65,Biskamp_Welter89}.   
 
Fractal, multifractal, and turbulent properties of photospheric 
active-region magnetic fields have been intensely studied in recent years.  
Fractality is traditionally investigated via the {\it fractal dimension},  
often inferred using box-counting techniques ({\it e.g.}, \opencite{Mandelbrot_83}).  
Box-counting is also used for multifractal studies in space and time  
({\it e.g.}, \opencite{Evertsz_Mandelbrot92}),  
involving also {\it generalized correlation dimensions}  
\cite{Georgoulis_etal95,Kluiving_Pasmanter96}.  
A commonly used multifractal method that 
does not require box counting is the calculation of the  
{\it multifractal structure function spectrum} \cite{Frisch_95}.  
Moreover, a practical method for 
quantifying turbulence is the calculation of the  
turbulent {\it power spectrum}, stemming from the original work of 
\inlinecite{Kolmogorov_41}.  
If the power spectrum shows a power law over a range 
of scales, perceived as the turbulent inertial range, 
its slope determines whether the inferred turbulence is Kolmogorov-like
(scaling index $\approx - 5/3$) or Kraichnan-like 
(scaling index $\approx - 3/2$) if either of these two applies. 

Multiple studies  
on fractality, multifractality, and turbulence in photospheric active-region  
magnetic fields have raised claims  
that flaring active regions exhibit distinct, distinguishable 
complexity. These works might lead to the  
impression that fractal, multifractal, or turbulent measures hold 
significant flare-predictive capability or, at least, they  
might be used to identify flaring active regions before they actually flare.  
To summarize some of these works,  
\inlinecite{Abramenko_etal03} suggested that a ``peak in the  
correlation length might be a trace of an avalanche of coronal  
reconnection events''. \inlinecite{McAteer_etal05} reported that  
``solar flare productivity exhibits an increase in 
both the frequency and GOES X-ray magnitude of flares from [active] regions  
with higher fractal dimension''. Further, \inlinecite{Abramenko_05}  
found that ``the magnitude of the power index at the stage of 
emergence of an active region ... reflects its future flare 
productivity when the magnetic configuration becomes well evolved'', while  
\inlinecite{Georgoulis_05} reported that ``the temporal evolution  
of the [inertial-range] scaling exponents in flaring active regions 
probably shows a distinct behavior a few hours prior to a flare''.  
More recently, \inlinecite{Conlon_etal08} worked on a  
sample of four active regions and  
reported evidence for a ``direct relationship between the  
multifractal properties of the flaring regions and their flaring rate'',  
while \inlinecite{Hewett_etal08}, reporting on  
``preliminary evidence of an inverse cascade in active region NOAA 10488''  
found a ``potential  
relationship between energy [power-spectrum] scaling and flare productivity''.  
Many of these works are also reviewed by \inlinecite{McAteer_etal10}.   
 
If the above findings are confirmed, they may well lead 
to notable improvements in our physical understanding of active 
regions and in highlighting possible differences between flaring  
(that is, hosting major flares) and  
non-flaring (that is, hosting only sub-flares) regions. In   
\inlinecite{Georgoulis_05} we studied three different  
scale-free and multiscale parameters,  
namely, the fractal dimension, the spectrum of generalized correlation  
dimensions, and the structure-function spectrum and its 
inertial-range exponents over a limited magnetogram sample of six active  
regions, three of them hosting at least one major flare (M- or X-class in  
the GOES X-ray 1--8 \AA$\;$  
flare classification scheme). In one case of a X-flaring active  
region -- NOAA active region (AR) 10030 with an X3 flare at the time of the  
observations -- we noticed a sharp preflare increase of the  
inertial-range exponent of the structure functions followed by a  
significant ($\approx 20$\% and much above uncertainties),  
permanent decrease after the flare.  
We suggested that this analysis should be repeated  
on a much larger sample of {\it both} flaring and non-flaring  
regions to determine whether this behavior was incidental.  
 
In this study we analyze a comprehensive sample of  
370  timeseries of active-region magnetograms, with each timeseries  
corresponding to a different active region. In this sample, 77 active  
regions hosted at least one M- or X-class flare during the 
observations and they are considered {\it flaring}  
(17 X-class flaring, 60 M-class flaring), while the remaining 293 active  
regions were not linked to major flares and are hence considered  
{\it non-flaring}. We calculate three of the most promising scale-free and  
multiscale measures on this data set, namely, the  
fractal dimension, the multifractal structure function spectrum, 
and the turbulent power spectrum. A detailed description of the data and  
techniques follows in Section~\ref{S-data_methods}.  
In Section~\ref{S-spres} we test the sensitivity of the calculated parameter
values on the spatial resolution of the studied magnetogram. 
A statistical analysis of the active-region sample is performed in  
Section~\ref{S-comp} while Section~\ref{S-conc} summarizes 
the study, discusses the results, and outlines our conclusions. 
\section{Design of the Study} 
\label{S-data_methods}  
\subsection{Magnetogram Data} 
\label{S-magdata}  
\bfig[t] 
\centerline{\includegraphics[width=.8\textwidth,clip=]{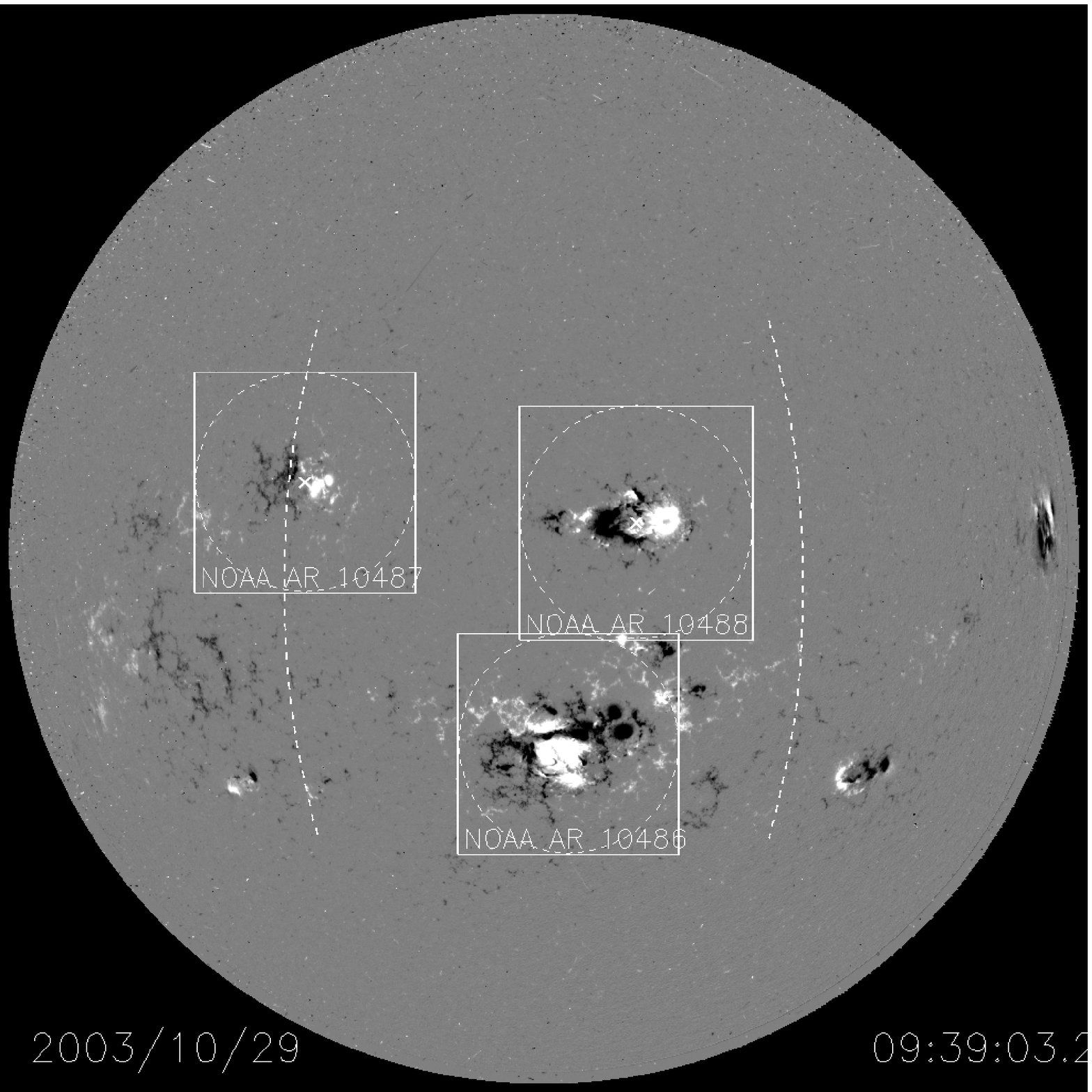}} 
\caption{Pictorial output of our ARIA, applied to a full-disk SOHO/MDI 
  magnetogram acquired on 29 October 2003. A $60^o$ meridional zone centered
  on the central meridian is indicated by the thick dashed brackets. 
  Three active regions fulfilled our 
  selection criteria, namely, NOAA ARs 10486, 10487, and 10488. The 
  portion of the disk found to include each region is shown by 
  the thin dashed circles - the actual portion extracted  
  for each region is shown by the circumscribed squares.  
  The NOAA labels for each region are provided automatically. 
  The flux-weighted centroids for 
  each active region are represented by white crosses.}  
\label{ARIA_example} 
\efig 
Our active-region sample has been constructed using data from the 
{\it Michelson-Doppler Imager} (MDI: \opencite{Scherrer_etal95}),  
onboard the {\it Solar and Heliospheric Observatory} (SOHO) mission. We 
acquired the entire MDI magnetogram archive from mid-1996 to 
late-2005. The archive consists of full-disk line-of-sight solar 
magnetograms taken at a 96-minute cadence with a linear pixel size of 
$\approx 1.98$ arcsec (a mean $\approx 1440$ km at solar disk center, 
depending on Sun--Earth distance). This 
analysis uses purely line-of-sight,  
Level 1.5 SOHO/MDI magnetic field 
measurements that are known \cite{Berger_Lites03} to underestimate 
sunspot and plage fields (more recent, Level 1.8.2 sensitivity 
corrections to the MDI full-disk magnetograms are not used  
in this study because our full-disk magnetogram  
dataset was constructed in 2007 
and the recalibrated magnetograms were posted in December 2008). 
Nonetheless, we have avoided applying 
any additional corrections to the data to avoid a possible impact on the 
morphological characteristics of the regions studied, since fractal 
and multifractal analysis highlights exactly these characteristics.  
To reduce the 
impact of projection effects acting on the magnetic field vector, we 
restrict our study to a $60^o$ longitudinal region centered on the central
solar meridian. Use of this zone introduces a systematic 
underestimation in the normal magnetic field component by a factor up 
to $\approx$ (1-cos($\theta$)) $\simeq 0.14$, or $14$\%, at the  
Equator, for a central meridian distance 
$\theta = 30^o$ approximated by the angular difference between the 
local normal and the line of sight for an observer at Earth. 
Conventionally, this underestimation factor is deemed 
tolerable when the line-of-sight field component is used as a proxy 
of the normal field component.  
Within this $60^o$ meridional zone we identified and extracted 
active regions using our automatic active-region identification 
algorithm (ARIA), detailed in  
\inlinecite{LaBonte_etal07} and in \inlinecite{Georgoulis_etal08}.  
Our ARIA extracts  
portions of the solar disk corresponding to active regions by means 
of pattern recognition in which the unit length is one 
supergranular diameter ($40$ arcsec). 
A typical example is shown in Figure \ref{ARIA_example}. 
An active region is chosen for further study if its flux-weighted 
centroid, shown by the white crosses in the selected regions of Figure  
\ref{ARIA_example}, falls within the above-mentioned $60^o$ meridional zone.  
Notice, for example, that NOAA AR 10487 is 
selected in Figure \ref{ARIA_example} because its flux-weighted 
centroid lies within the above zone; parts of it, however, 
extend beyond this area. 
 
Besides the automatic active-region selection process,  
each selected magnetogram (out of a total of 17,733) was manually  
examined to exclude portions of other 
active regions that might intrude in the field of view. For 
example, NOAA AR 10486 in Figure \ref{ARIA_example} is included in its 
selection circle, but the square circumscribed on this circle 
crops sizable parts of NOAA ARs 10489 and 10491 in its  
northwestern edge. These parts have been excluded in the 
subsequent analysis. Generally our ARIA performs quite well in 
distinguishing active regions but few incidences such as the 
above have been noted, especially in cases of densely populated  
active-region complexes, or ``nests'',  
such as the one believed to have occurred during the 
October-November ``Halloween'' 2003 
period \cite{Zhou_etal07}. For our analysis, 
ARIA uses a maximum tolerated magnetic-flux imbalance of $50\%$ in a 
given active region and a minimum active-region linear size of one 
supergranular diameter. For each of the 370 selected active regions we 
created a timeseries consisting of up to $\approx 60$ magnetograms taken 
every 96 minutes corresponding to the approximately 
four-day period needed for each active region 
to traverse the $60^o$-analysis zone.       
 
To document the major flare history for each active region we browsed 
{\it i)} NOAA's GOES X-ray archive and {\it ii)} the Yohkoh/HXT flare  
catalog (available online, at  
\url{http://gedas22.stelab.nagoya-u.ac.jp/HXT/catalogue/}). 
From the total of 370 active regions, 77 were unambiguously found to 
have hosted at least one M-class or X-class flare while within  
$\pm 30^o$ of the central meridian, with a total of 24 X-class 
flares and 87 M-class flares. Our active-region sample roughly covers 
Solar Cycle 23, as shown in Figure \ref{ARdis}. The solar cycle is 
represented by a 5-point running mean of the monthly-averaged sunspot number
obtained by the Solar Influences Data Analysis Center (SIDC) of the Royal
Observatory of Belgium.
\bfig[t] 
\centerline{\includegraphics[width=1.\textwidth,clip=]{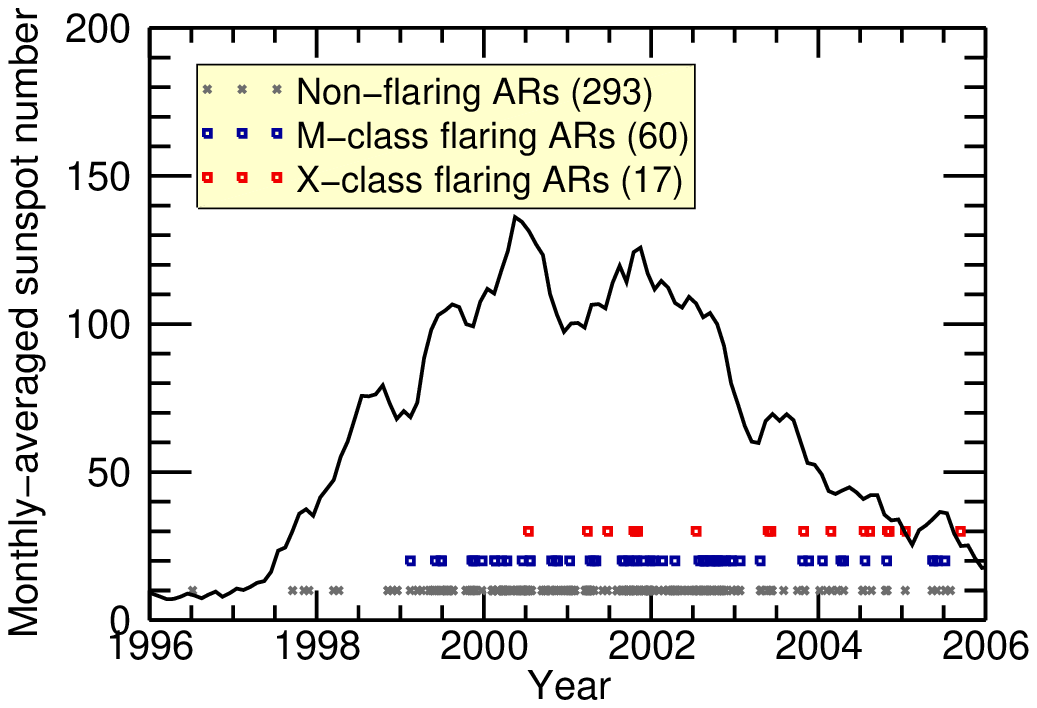}} 
\caption{Temporal distribution of our sample of 
  370 active regions over Solar Cycle 
  23. Shown are the median observation times (color symbols) of each 
  region together with the monthly-averaged sunspot number (curve).}
\label{ARdis} 
\efig

\bfig[t] 
\centerline{\includegraphics[width=0.8\textwidth,clip=]{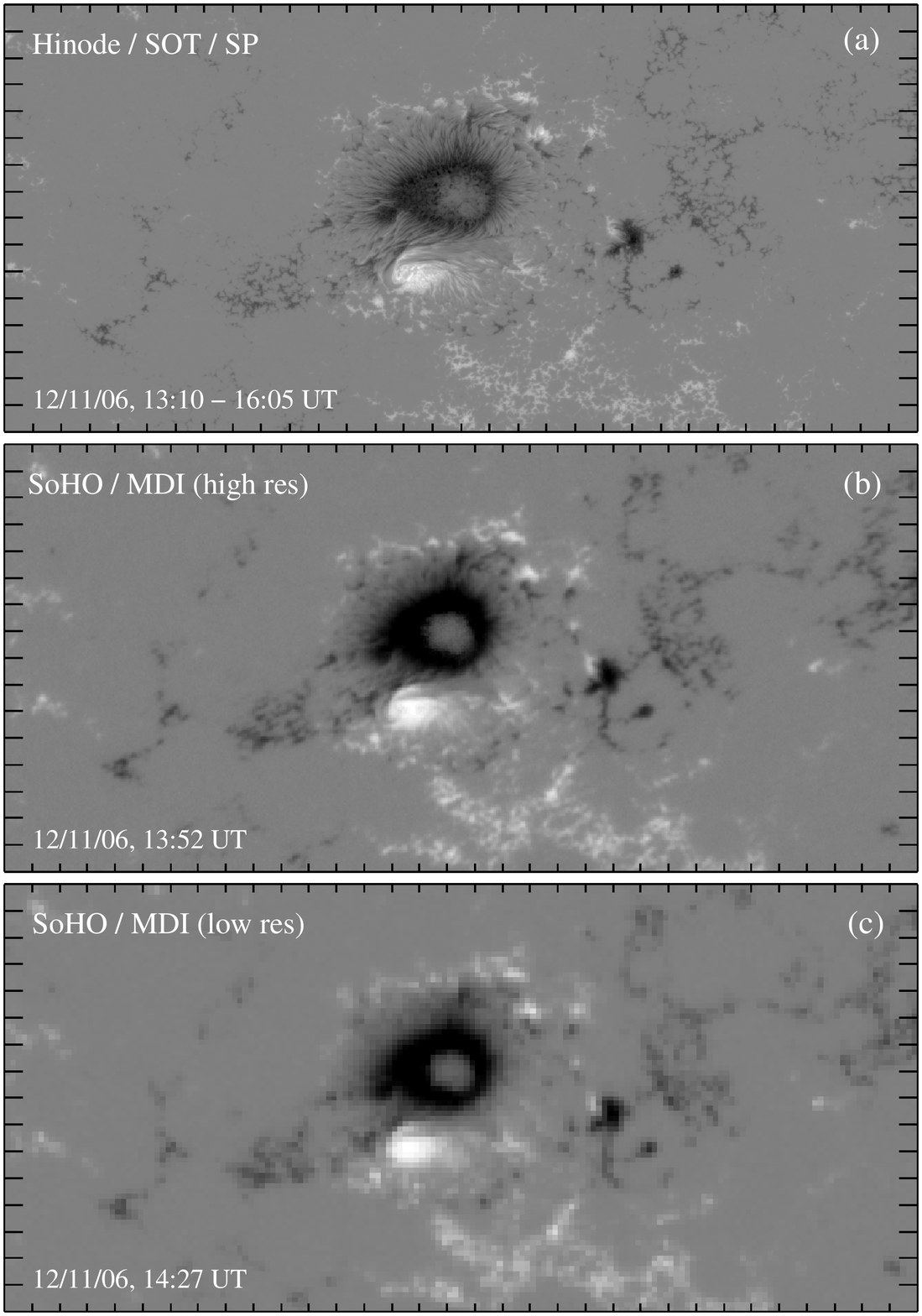}} 
\caption{Nearly simultaneous, coaligned magnetograms of NOAA AR  
  10930, acquired on 11 December 2006: observations are from (a)    
  Hinode's SOT/SP, (b) SOHO/MDI 
  high-resolution, partial disk magnetograph, and (c) SOHO/MDI 
  full-disk magnetograph. The linear pixel sizes are 
  $0.158$ arcsec, $0.605$ arcsec, and $1.98$ arcsec for 
  (a), (b), and (c), respectively. 
  Shown is the  line-of-sight magnetic field component saturated at 
  $\pm 2.5\;kG$ (a), $\pm 1$ kG (b), and $\pm 1.8$ kG (c).  
  Tic mark separation in all images is $10$ arcsec.} 
\label{AR930} 
\efig 
In addition to the above SOHO/MDI sample, our analysis includes 
three nearly simultaneous magnetograms of NOAA 
AR 10930, observed on 11 December 2006. The line-of-sight components 
of these magnetograms are depicted in Figure \ref{AR930}. Figure 
\ref{AR930}a shows the Level 1D  
magnetogram (preferred 
over Level 2 data in order to better qualify for comparison with 
SOHO/MDI Level 1.5 data) acquired by the Spectropolarimeter  
(SP: \opencite{Lites_etal01}) of the {\it Solar Optical Telescope} (SOT) 
onboard the {\it Hinode} 
satellite and has a very high spatial resolution ($\approx 
0.32$ arcsec) with a linear pixel size of $\approx 0.158$ arcsec. Figure 
\ref{AR930}b shows the respective magnetogram taken by the SOHO/MDI 
high-resolution, partial-disk magnetograph, with a coarser linear 
pixel size of $0.605$ arcsec. Figure \ref{AR930}c shows the Level 1.5 
SOHO/MDI magnetogram extracted by a full-disk measurement with a much 
coarser linear pixel size of $1.98$ arcsec. The three magnetograms 
have been initially coaligned by means of the pointing information 
provided separately for each. To further correct and  
deal with small pointing inconsistencies, coalignment has been 
completed by a rigid displacement (translation) over the E--W and the 
N--S axes. Displacements are determined by the peak of the
  cross-correlation function between a given pair of images, with   
cross-correlation functions inferred by means of fast Fourier transforms.
 
The three distinctly 
different spatial resolutions of the magnetograms of Figure 
\ref{AR930} will be useful when testing the sensitivity of our   
scale-free and multiscale parameters  
to varying spatial resolution (Section~\ref{S-spres}).   
\subsection{Scale-free and Multiscale Techniques} 
\label{S-tech} 
The first parameter that we calculate is the scale-free, two-dimensional  
{\bf fractal dimension} $[D_0]$ of the 
active-region magnetograms. To calculate $D_0$ we cover the 
magnetogram field-of-view with a rectangular grid 
consisting of square boxes with linear size $[\lambda]$ and area 
$\lambda \times \lambda$. Assuming that the field of view is a square 
with linear size $L$ and area $L \times L$, the number of boxes needed 
to cover it is $(L/\lambda)^2$. Each of the boxes will have a 
dimensionless area $\varepsilon ^2$, where  
$\varepsilon = \lambda / L$. Of the total $(L/\lambda)^2$  
boxes we count 
those that include part of the boundary of a strong-field magnetic 
configuration (see below for the adopted strong-field definition). 
Then, varying the box size 
$\lambda$ or, equivalently, the dimensionless size $\varepsilon$, we 
obtain different numbers $[N(\varepsilon)]$ of information-carrying 
boxes. Correlating the various numbers $N(\varepsilon)$ with the 
respective box sizes $\varepsilon$, we obtain the scaling 
relation  
\beq 
N(\varepsilon) \propto (1/\varepsilon)^{D_0}\;\;. 
\label{D0} 
\eeq 
For a non-fractal, Euclidean structure embedded on a plane we have  
$N(\varepsilon) = (L/\lambda)^2 = (1/\varepsilon)^2$, so $D_0=2$.  
If $D_0 <2$, we have fractal structures with a scale-free, incomplete 
filling of the field of view. The stronger the departure of $D_0$ from 
its Euclidean value of 2, the finer the structure exhibited by the 
studied magnetic configuration. For $D_0 \le 1$ in a two-dimensional 
fractal, the structures are typically scattered into a scale-free 
hierarchy of small ``islands', resembling what is known as  
{\it fractal dust} ({\it e.g.} \opencite{Schroeder_91}).  
 
We infer the fractal dimension $D_0$ by a least-squares best fit of  
the scaling relation of Equation (\ref{D0}). The uncertainty 
associated with the value of $D_0$ is equal to  
the uncertainty of the regression fit. To guarantee a reliable 
inference of $D_0$, we demand that the dynamical range represented by the 
least-squares fit exceeds one order of magnitude. A very small fraction
  of magnetograms of non-flaring active regions ($\simeq 0.6$\%, or 
  111 magnetograms) happen not to comply with this requirement because of
  the regions' simplicity and scattered configurations; these magnetograms
  have been excluded from the analysis. Nonetheless, each of the 370 active
  regions of our sample fulfills the requirement with at least one magnetogram.
 
\inlinecite{McAteer_etal05}, relying on a substantial data set of  
$\approx 10^4$ active regions, first reported that flaring regions have 
fractal dimensions $D_0 \ge 1.2$. Their finding was statistical, of 
course, meaning that the $D_0 \ge 1.2$ condition should be viewed as a 
necessary, but not sufficient, condition for major flare 
productivity. They also concluded, and we test their result here,  
that intensely flaring active regions showed statistically  
higher fractal dimensions.  
 
To better compare with the results of \inlinecite{McAteer_etal05}, we 
follow their criterion when outlining the boundaries of active regions: 
first, we use a threshold of $50$ G in the strength of the 
line-of-sight field component in order to define the outer contours of 
strong-field magnetic patches. Then we impose a lower limit of 20 
pixels for the length of each contour, thus rejecting  
very small patches that could as easily belong to the quiet Sun. As an 
additional condition, we impose a lower flux limit of $10^{20}$ Mx for 
each patch. For the MDI low-resolution 
data this is nearly equivalent to saying that at least 100 pixels 
within the patch should have a line-of-sight field strength of at 
least $50$ G, which is our threshold. Of course, a selected patch can 
contain fewer than 100 (but more than 20) pixels but with larger field 
strength in order to satisfy the flux condition.   
 
The second parameter that we calculate is the {\bf multifractal structure 
  function spectrum} \cite{Frisch_95}. The spectrum is given by  
\beq 
S_q(r) = \langle |\Phi(\mbf{x}+\mbf{r}) - \Phi(\mbf{x})|^q \rangle 
\label{Sq_r1} 
\eeq 
and it does not rely on box-counting or thresholding, contrary to
  $D_0$. Instead, on the magnetic-flux 
distribution $[\Phi (\mbf{x})]$ of the active-region photosphere we  
define a {\it displacement} vector $[\mbf{r}]$, also called the separation 
vector, and calculate the variation of the flux at this 
displacement. The variation is then raised to the power $q$, where $q$ 
is a real, preferably positive number called the {\it selector}.  
Spatial averaging ($\langle \rangle$) of the structure function over 
$\mbf{x}$ and all possible orientations of $\mbf{r}$ gives rise to a 
unique, positive value $S_q(r)$ of the structure function for a given 
pair $(r=|\mbf{r}|,q)$. The resulting spectrum involves a range of 
$r$-values and a fixed value of $q$; different spectra are obtained 
for different $q$-values. 
 
The multifractal structure function is designed to highlight the 
intermittency present in a magnetic-flux distribution  
(\opencite{Abramenko_etal02}; \citeyear{Abramenko_etal03}).  
In the case of a multifractal, intermittent flux 
distribution, the structure function $[S_q(r)]$ exhibits a power law  
\beq 
S_q(r) \propto r^{\zeta(q)} 
\label{Sq_r2} 
\eeq 
within a range of displacements, often referred to as the turbulent 
inertial range. The upper and lower extremes of the $r$-range correspond 
to, respectively,  
the maximum size of structures entering the inertial range and the 
scale over which ideal cascading of energy to smaller scales breaks 
down by dissipative effects. Higher values of the inertial-range 
scaling index $\zeta (q)$ indicate a higher degree of intermittency, 
with $\zeta (q) = q/3$ implying absence of intermittency.  
 
\inlinecite{Abramenko_etal08} studied the structure function spectrum 
and implemented an additional suggestion by \inlinecite{Frisch_95} to examine 
the {\it flatness function} $f = S_6(r)/S_3(r)^3 \propto r^{-\delta}$, 
where $\delta$ is viewed as the {\it intermittency index} with higher 
values implying a higher degree of intermittency. They  
concluded that photospheric and coronal magnetic fields 
are both intermittent, with the intermittency in the photosphere 
preceding that in the corona and the corona responding to photospheric 
increases of intermittency. In their analysis, they used data from the 
same active region we show in Figure \ref{AR930}, namely, 
NOAA AR 10930, observed by {\it Hinode}/SOT/SP and SOHO/MDI high-resolution 
magnetographs, although not on the same day with the data used in this 
study.  
 
\inlinecite{Georgoulis_05} used the structure function 
$S_q(r)$ of Equation (\ref{Sq_r2}) 
and the inertial-range scaling exponent $\zeta (q)$ to show 
that the photospheric magnetic fields of the six active regions of  
the study were indeed multifractal and intermittent, departing 
strongly from $\zeta (q) = q/3$. More importantly,  
\inlinecite{Georgoulis_05} presented  
an example of a X3 flare that occurred in NOAA AR 10030, 
where the intermittency peaked $\approx 1-2$ hours prior to the flare and 
decreased sharply after the flare. The decrease was permanent and 
$\zeta (q)$ continued decreasing $\approx 1.5$ hours later, when 
an M1.8 flare also occurred in the active region (Figure 7 of 
\opencite{Georgoulis_05}).  
The change in the degree of intermittency was best seen for a 
selector $q \in (3,3.5)$. Here we investigate whether this 
distinct behavior, seen in only {\it one} example,   
is part of a systematic tendency.  
To this purpose we study the temporal 
evolution of the scaling index $\zeta (q=3)$, inferred by a 
least-squares best fit of the scaling relation of Equation 
(\ref{Sq_r2}). The uncertainty of the regression fit is treated as the 
uncertainty of the value of $\zeta (3)$. For the uncertainty of the 
$\zeta (3)$-differences we simply propagate the uncertainties of the 
two $\zeta (3)$-values that make these differences.  
 
The third parameter we study is the {\bf turbulent power spectrum} 
$[E(k)]$. In the case of a turbulent flux system, there exists an inertial 
range of wavenumbers $k$ reflected on a power-law form for $E(k)$:  
\beq 
E(k) \propto k ^{- \alpha}\;\;, 
\label{tps} 
\eeq 
where $\alpha$ is the inertial-range exponent. We demand  
at least one order of magnitude as the dynamical 
range of the least-squares best fit used to infer $\alpha$ and we 
attribute the uncertainty of the fit to the  
uncertainty of $\alpha$. Like $S_q(r)$ and unlike $D_0$, no box-counting 
or thresholding is required to infer $\alpha$.

\inlinecite{Abramenko_05}  
suggested that the $k$-range  
corresponding to length scales $r \in (3,10)$ Mm should be scaled 
according to the representative $\alpha$-value for high-resolution MDI 
magnetograms. Although we use MDI low-resolution 
data for this analysis, we follow 
\citeauthor{Abramenko_05}'s suggestion and make sure that the 
$(3,10)$ Mm scale range is included in the power-law fit. We  
also attempt to test the conclusion of \inlinecite{Abramenko_05}, 
relying on a sample of 16 high-resolution MDI active-region 
magnetograms, that the inertial-range exponent $\alpha$ reflects 
the future flare productivity of an active region, with larger 
$\alpha$-values implying a higher flare index (Figure 8 of  
\opencite{Abramenko_05}). It is also important to mention here  
\citeauthor{Abramenko_05}'s (\citeyear{Abramenko_05}) conclusion  
that the scaling index $\alpha$ is not particularly useful 
for the prediction of imminent flares in the studied active regions.  
 
\bfig[t] 
\centerline{\includegraphics[width=1.\textwidth,clip=]{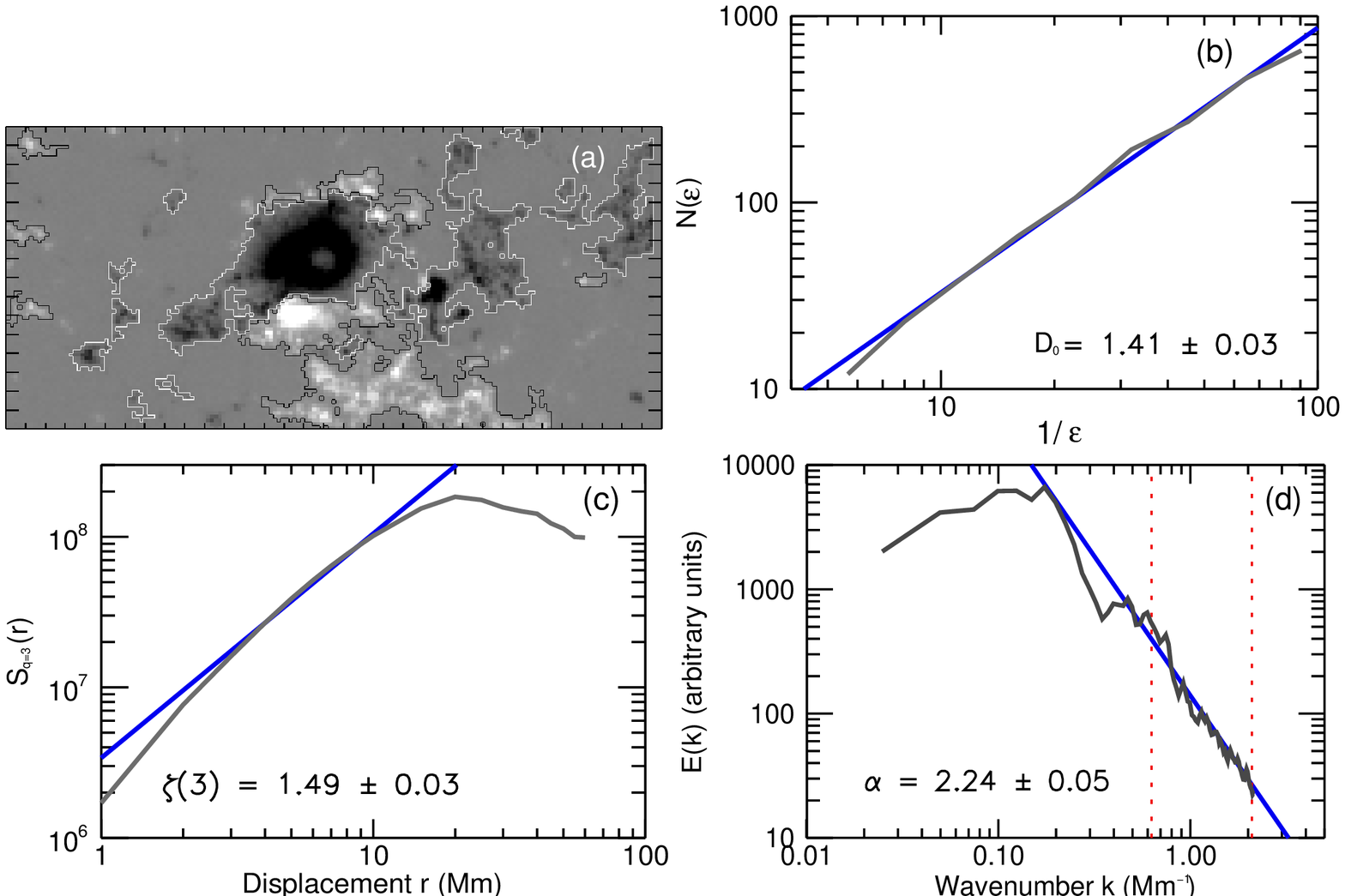}} 
\caption{Typical calculation of the scale-free and multiscale
  parameters that will be used to quantify the
  complexity of solar active regions. The example refers to the
  low-resolution SOHO/MDI magnetogram of Figure \ref{AR930}c, shown in
  (a). The magnetogram is saturated at $\pm 1$ kG. Contours indicate
  the areas where the line-of-sight field strength exceeds the
  threshold of  
  50 G. Tic mark separation is 10 arcsec. Also shown are (b) the
  scaling relation of Equation (\ref{D0}) yielding the fractal
  dimension $D_0$, (c) the structure function $S_{q=3}(r)$ of Equation
  (\ref{Sq_r1}), yielding the inertial-range scaling index $\zeta
  (3)$, and (d) the turbulent power spectrum of Equation (\ref{tps}),
  yielding the inertial-range scaling exponent $\alpha$. The two
  dotted lines indicate the wavenumbers of the desired
  length scales of 10 Mm ($k <1$) and 3 Mm ($k >1$). Gray curves in b,
  c, and d correspond to $N(\varepsilon)$, $S_{q=3}(r)$, and $E(k)$,
  respectively, while the blue lines are the respective 
  least-squares best fits.}
\label{example} 
\efig
A typical calculation of the fractal dimension $D_0$, the
inertial-range scaling index $\zeta(3)$, and the scaling 
exponent $\alpha$ of the turbulent power spectrum for the
low-resolution MDI magnetogram of Figure \ref{AR930}c is depicted in
Figure \ref{example}.

With the exception of the turbulent power 
spectrum, \inlinecite{Georgoulis_05} used all of the 
above methods, including the spectrum of multifractal generalized 
correlation dimensions. The latter showed that the six active regions of 
the sample had clearly multifractal photospheric magnetic fields. The 
method is not used in this study, however,  
because in \inlinecite{Georgoulis_05} it 
clearly failed to distinguish between  
flaring and non-flaring regions even 
for the handful of active regions studied.  
 
The objective of this work is to test the flare-predictive capability 
of the above -- reported to be promising -- parameters. 
This capability will be tested by comparison 
with the capability of a standard, traditional parameter reflecting 
the size of active regions, namely, the {\bf unsigned magnetic flux}  
\beq 
\Phi _{tot} = \int _S |B_n| \mathrm{d}S\;\;, 
\label{ftot} 
\eeq 
where $B_n$ is the normal magnetic field component (approximated by 
the line-of-sight field component near disk center) over the magnetograms'  
field of 
view $[S]$. Flaring regions are statistically more flux-massive 
than non-flaring ones. As a result, the unsigned flux is considered a 
``standard'' flare forecasting criterion that, however, is not without 
limitations ({\it e.g.}, \opencite{Leka_Barnes03}), especially when it comes 
to flux-massive, but quiescent, active regions.  
For a parameter to be characterized as 
having a significant predictive  
capability, it has to perform better than the unsigned flux.  
\section{Dependence of Scale-Free and Multi-Scale Parameters on the Spatial Resolution}
\label{S-spres} 
Prior to analyzing our extensive MDI dataset we perform a sensitivity test of
the parameters introduced in Section~\ref{S-tech} to the spatial resolution of
the studied magnetogram. Ideally, a parameter reliable enough to  
distinguish flaring from non-flaring active regions should be fairly 
insensitive to varying spatial resolution. 
If this were not the case, then one should at least be able 
to model the parameter's variations with 
changing resolution. If these conditions are not fulfilled, likely  
threshold values of the parameter that one may use to identify  
flaring regions before they flare are resolution-dependent and, 
as a result, instrument-dependent.
 
\begin{table}[t] 
\caption{Values and uncertainties (in parentheses) of the unsigned magnetic
  flux $\Phi_{\mathrm{tot}}$, the fractal dimension $D_0$, the inertial-range 
  scaling exponent $\zeta (3)$, and the power-spectrum index $\alpha$ for the
  three nearly simultaneous and coaligned magnetograms of NOAA AR 10930 
  acquired on 11 December 2006 (Figure \ref{AR930}).} 
\label{tab1} 
\begin{tabular}{ccccccc} 
\hline 
Ref.            & Observation & Pixel size & $\Phi_{\mathrm{tot}}$ & & & \\
(Fig.)          & Time (UT)   & (arcsec)   & ($\times 10^{22}$ Mx) & $D_0$ & $\zeta (3)$ & $\alpha$\\                 
\hline
\ref{AR930}a    & 13:10 -- 16:05 & 0.158 & 2.85 & 1.54 (0.04) & 1.35 (0.03) & 3.00 (0.02)\\
\ref{AR930}b    & 13:52:01       & 0.605 & 2.12 & 1.43 (0.02) & 1.67 (0.04) & 3.32 (0.04)\\
\ref{AR930}c    & 14:27:01       & 1.980 & 3.60 & 1.41 (0.03) & 1.49 (0.03) & 2.24 (0.05)\\
\hline 
\end{tabular} 
\end{table} 
We use the three magnetograms of NOAA AR 10930 shown in Figure 
\ref{AR930} to test the parameter values on data with varying spatial
resolution. The results are provided in Table \ref{tab1} for the parameters 
$\Phi _{tot}$ [Equation (\ref{ftot})], $D_0$ [Equation (\ref{D0})], 
$\zeta (q=3)$ [Equation (\ref{Sq_r2})], and $\alpha$ [Equation (\ref{tps})]. 
Our findings can be summarized as follows: 
\ben 
\item[{\it i)}] Despite coalignment and near simultaneity, 
  the unsigned magnetic flux  $\Phi _{\mathrm{tot}}$ shows distinct
  differences for the three different magnetograms: the MDI low-resolution
  magnetogram (Figure \ref{AR930}c) shows $\approx 25$\% larger flux than
  the SOT/SP magnetogram (Figure \ref{AR930}a), while the MDI high-resolution
  magnetogram (Figure \ref{AR930}b) shows $\approx 28$\% less flux than the
  SOT/SP magnetogram. Some additional testing, not shown here, has been
  performed in an attempt to explain these large differences. In particular,
  attempting to degrade the 
  highest-resolution magnetograms (SOT/SP and high-resolution MDI) to simulate
  situations of a larger pixel size (2 arcsec and more) we find a very weak
  decreasing trend for the unsigned flux that reaches up to $\approx 3.5$\%
  between the highest- (original) and the lowest- (degraded) resolution
  magnetogram. Likely, therefore, the $\approx 30$\% flux difference is not
  due to the different spatial resolution. Similar results are obtained
  when the spatial resolution of the SOT/SP and high-resolution MDI data 
  is decreased by resampling. 
  We cannot be certain about the source(s) of the discrepancy 
  but, given that we have used SP and MDI data of similar 
  calibration levels, one might attribute this discrepancy to differences in 
  the observations, processing, and calibration between the SP and the two MDI
  magnetographs. Discussing those differences exceeds the scope of this work. 
\item[{\it ii)}] The values of the scale-free fractal dimension $D_0$ are 
  fairly consistent, despite the widely different spatial resolution and the
  different instruments. One
  notices a significant (i.e., beyond error bars) 
  decreasing tendency for decreasing resolution 
  but the overall decrease for the $\approx 13$-fold difference in 
  resolution between the SOT/SP and the low-resolution MDI data is only
  $\approx 8.5$\%. This is probably because the fractal dimension 
  qualitatively highlights the morphological complexity of the studied 
  self-similar structure that is being reflected adequately on seeing-free 
  (SOT/SP and MDI) magnetograms largely regardless of spatial 
  resolution and magnetic flux content. 
\item[{\it iii)}] The values of the multiscale inertial-range scaling exponent 
  $\zeta (3)$ and the power-spectrum scaling index $\alpha$ appear strongly
  dependent on the spatial resolution and/or other instrumental
  characteristics. The variation of both indices is not even monotonic,
  with their peak values corresponding to the intermediate case of the
  high-resolution MDI magnetogram. One might speculate that this happens 
  because different spatial resolution changes quantitatively, but perhaps not
  qualitatively, the multiscale character of the data. In this sense, despite
  different parameter values, all three magnetograms of
  NOAA AR 10930 show significant intermittency ($\zeta (3) \gg 1$, with lack of
  intermittency reflected on $\zeta (3) =1$) and non-Kraichnan/non-Kolmogorov 
  turbulence ($\alpha \gg 3/2$ and $5/3$, respectively). 
\een 
  
The susceptibility of the multiscale parameters $\zeta (3)$ and 
$\alpha$, but not of the scale-free parameter $D_0$, to the 
spatial resolution implies caution when utilizing multiscale parameters to 
distinguish flaring from non-flaring active regions. At the very least,
quantitative results in this case should not be generalized to 
different data sets. We therefore stress that the results
described in the next Sections for the multiscale parameters correspond 
{\it exclusively} to the MDI 
full-disk spatial resolution of $\approx 2$ arcsec.  
\section{Comparison of Parameters for Flaring and Non-Flaring Active Regions} 
\label{S-comp} 
To determine whether any of the fractal or 
multifractal parameters $D_0$, $\zeta (3)$, and $\alpha$, including 
$\Phi_{\mathrm{tot}}$ as a reference, can distinguish flaring from non-flaring 
active regions, we perform two tests: a more stringent one, that 
compares the {\it preflare} (96 minutes in advance, 
at most, per the cadence of the full-disk MDI magnetograms) 
values of the parameters for flaring active 
regions to the {\it peak} values of the parameters for non-flaring  
regions, and a more liberal one, that compares the peak values of the 
parameters for both flaring and non-flaring regions.  
Finding a distinguishing pattern in the first test would mean that the 
studied parameter may have a short-term predictive capability.  
If the first test fails but the second test gives 
some distinguishing patterns, the studied parameter 
provides clues about the expected flare productivity 
of an active region, without necessarily implying  
when flares will occur. This claim is found in multiple 
instances in the literature  
({\it e.g.}, \opencite{McAteer_etal05} for $D_0$; \opencite{Abramenko_05} 
and \opencite{Hewett_etal08} for $\alpha$; \opencite{Conlon_etal08} 
for other multifractal parameters).
\subsection{Preflare vs. Peak Non-Flaring Parameter Values}
\label{S-prepeak}
\bfig[t] 
\centerline{\includegraphics[width=1.\textwidth,clip=]{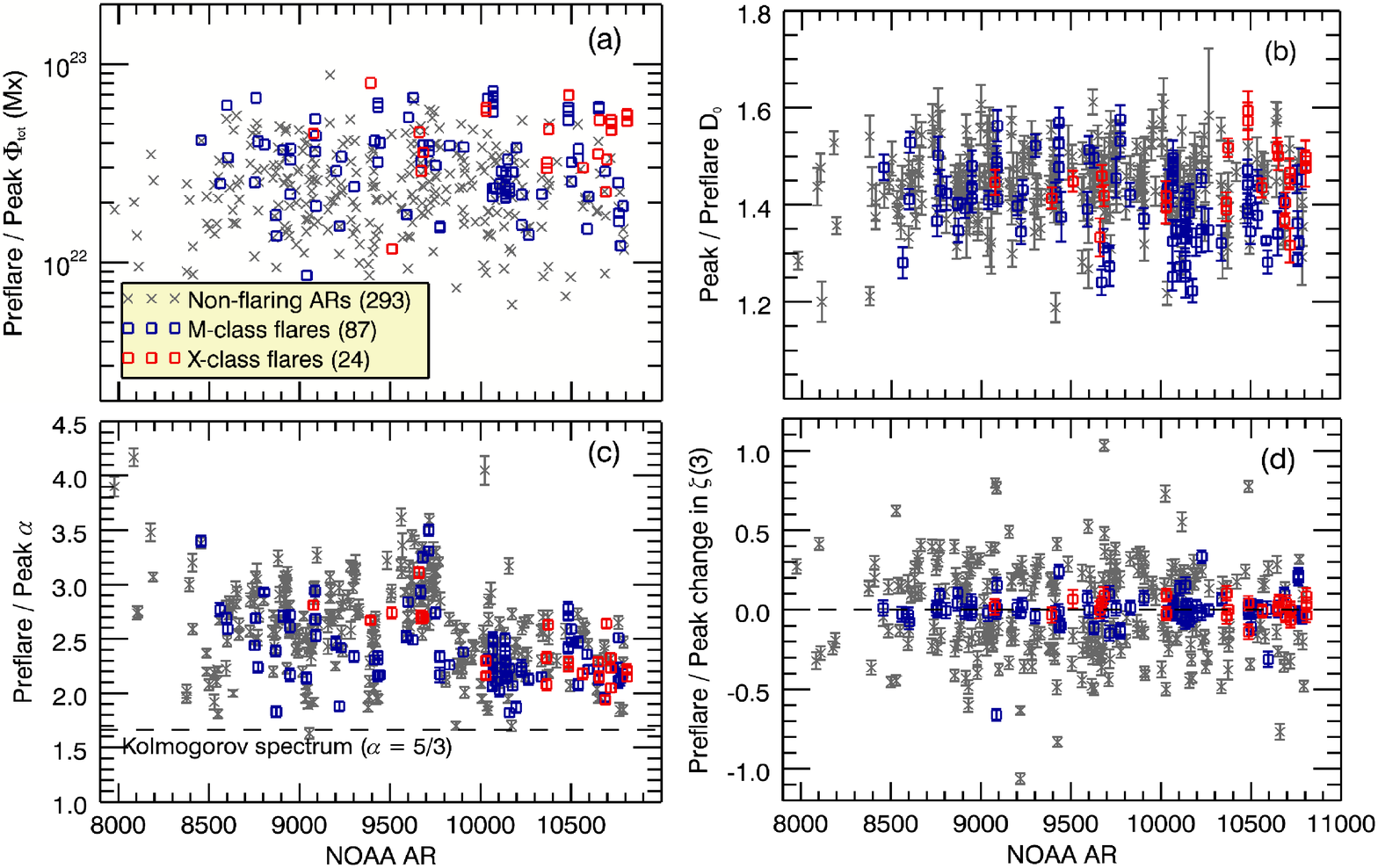}} 
\caption{Comparison between the preflare values of scale-free and multiscale 
  parameters of flaring active regions and the respective peak values of 
  non-flaring regions. Shown are (a) the unsigned magnetic flux $\Phi _{tot}$, 
  (b) the fractal dimension $D_0$, (c) the turbulent power-spectrum index
  $\alpha$, and (d) the change in the inertial-range scaling exponent $\zeta (3)$.} 
\label{pre_max} 
\efig 
Figure \ref{pre_max} depicts the preflare values of $\Phi_{\mathrm{tot}}$ 
(Figure \ref{pre_max}a), $D_0$ (Figure \ref{pre_max}b), and  
$\alpha$ (Figure \ref{pre_max}c). It also provides the change between 
the preflare and the postflare values of $\zeta (3)$ (Figure 
\ref{pre_max}d) for flaring regions, compared with the peak $\zeta 
(3)$-change for non-flaring ones. Flaring regions are 
divided into ``M-flaring'' (regions that have given at least one 
M-class, but not a X-class, flare; blue squares) and ``X-flaring'' 
(regions that have given at least one X-class flare; red squares).  
Since peaks of the studied parameter timeseries are taken, each 
active-region timeseries has been inspected separately to remove spurious 
effects in the parameters' evolution. The most prominent source of 
these effects is contamination due to flare emission, in the case of large 
white-light flares, with temporary instrumental or data problems playing 
a secondary role. In case spurious effects are detected, the affected 
parameter value is replaced by the value interpolated for a given time. 
 
Inspecting Figure \ref{pre_max}, we first notice that only the  
$\Phi_{\mathrm{tot}}$-values of the flaring regions (Figure \ref{pre_max}a) 
show some tendency to occupy the upper $\Phi_{\mathrm{tot}}$-range in 
the plot. The preflare values of the fractal dimension $D_0$  
(Figure \ref{pre_max}b) appear to be more or less uniformly 
distributed between $\approx 1.2$ and $\approx 1.8$, with a mean value of 
$\approx 1.41$ and a standard deviation of $\approx 0.08$. For  
non-flaring regions, the peak fractal dimension has a mean of   
$\approx 1.44$, with a standard deviation $\approx 0.14$.  
Qualitatively, we 
reproduce the result of \inlinecite{McAteer_etal05} who found  
$D_0 \gtrsim 1.2$ for flaring regions, but we also find that  
$D_0 \gtrsim 1.2$ for all active regions in our sample.  
 
A similar behavior is seen when the scaling index $\alpha$ 
of the turbulent power spectrum is examined (Figure 
\ref{pre_max}c). Indeed, the preflare $\alpha$-values show a mean 
$\approx 2.4$ and a standard deviation $\approx 0.32$, while the peak 
$\alpha$-values for the non-flaring regions have a mean $\approx 2.60$ 
with a standard deviation $\approx 0.42$. No active region shows a 
$\alpha$-value smaller than the Kolmogorov index of $5/3$ and very 
few regions, among them three M-flaring ones, show 
$\alpha <2$. We conclude that a strong 
departure from a Kolmogorov turbulent spectrum is {\it not} a 
characteristic of {\it some} (flaring) active regions but one of 
{\it most} active regions. Given the $\alpha$-dependence on spatial
resolution, however (Section \ref{S-spres}), we cannot be certain about the
``true'' $\alpha$-value. 
 
For the multifractal inertial-range scaling exponent $\zeta (3)$ we 
have compared the change in values between the preflare and the 
postflare phase with the peak change in values for non-flaring active 
regions (Figure \ref{pre_max}d). This was chosen because in 
\inlinecite{Georgoulis_05} we inferred a significant, permanent 
decrease in $\zeta (3)$ from preflare to postflare in a single flaring 
region. Unfortunately, this feature does not survive here, where more 
comprehensive statistics are involved: for the 111 major flares 
included in our sample (24 X-class, 87 M-class) the host active 
regions show a preflare / postflare increase in $\zeta (3)$ in 59 
cases, with a decrease between the preflare and postflare  
$\zeta (3)$-values inferred in the remaining 52 cases.  
Clearly, the peak $\zeta (3)$ changes do not correspond to flaring 
active regions with the exception of the two M-flaring regions NOAA AR 
9087 and 10596, that show two of the sharpest $\zeta (3)$-decreases in 
the postflare phase, with amplitudes 0.66 and 0.31, respectively. If 
these two regions were studied in isolation and one ignored the 
$\zeta$-dependence on the spatial resolution, then one might have 
concluded that large flares indeed relate to sharp decreases in  
$\zeta (3)$, meaning a decrease of the degree of the photospheric  
intermittency in the active regions after the flare. Given our 
spatial-resolution test and the large active-region sample, however, 
such a conclusion is unjustified. 
 
The apparent failure of the first, stringent test comparing the preflare 
values of the studied parameters for flaring regions with the respective peak 
values for non-flaring ones indicates that none of the scale-free and 
multiscale parameters shows any  
notable short-term flare prediction capability or, at 
least, any better predictive capability than the conventional unsigned 
magnetic flux. 
\subsection{Peak Flaring vs. Peak Non-Flaring Parameter Values}
\label{S-allpeak}
We now perform the second test by comparing the 
peak values of the studied parameters for both flaring and for non-flaring 
regions. The results of this test are shown in Figure \ref{all_max}. 

\bfig[t] 
\centerline{\includegraphics[width=1.\textwidth,clip=]{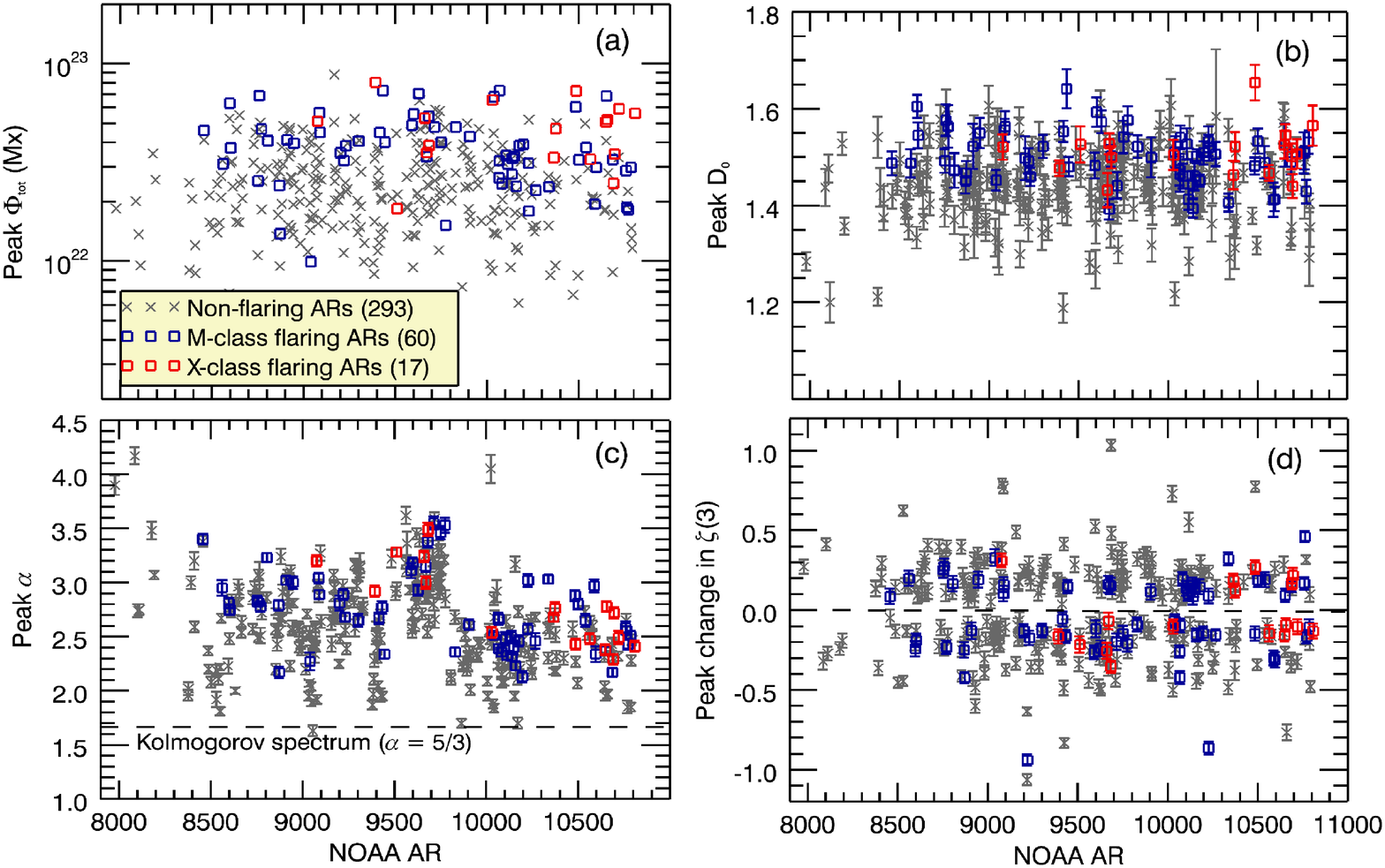}} 
\caption{Same as Figure \ref{pre_max}, but showing the peak, rather than the 
  preflare, values of the examined parameters for the flaring active regions.} 
\label{all_max} 
\efig 
Figure \ref{all_max}a depicts the results for the unsigned magnetic flux  
$\Phi_{\mathrm{tot}}$. It shows more clearly than Figure \ref{pre_max}a  
that the peak $\Phi_{\mathrm{tot}}$-values tend to 
occupy the upper part of the $\Phi_{\mathrm{tot}}$-range. Somewhat 
more tell-tale is the difference between Figures \ref{pre_max}b and 
\ref{all_max}b in terms of the preflare and the peak fractal 
dimension, respectively: the peak $D_0$-values for flaring regions in 
Figure \ref{all_max}b also tend to occupy the upper part of the 
$D_0$-range, meaning that flaring regions statistically tend to have 
higher fractal dimension, as \inlinecite{McAteer_etal05}  
previously reported.     
 
Regarding the indices $\alpha$ of the turbulent power spectrum 
(Figure \ref{all_max}c), there is no distinct difference with 
Figure \ref{pre_max}c. 
The peak $\alpha$-values for flaring regions also tend to 
occupy the higher $\alpha$-range but with larger dispersion than the 
fractal dimension $D_0$. This being said, the active regions with the 
highest peak $\alpha$-values in our sample happen to be non-flaring 
ones.  
 
When the peak $\zeta (3)$ change for both flaring and non-flaring 
regions is considered (Figure \ref{all_max}d) we find that flaring 
regions show a similar pattern with non-flaring ones. Only for the 
X-flaring regions there seems to be a weak statistical preference for 
stronger $\zeta (3)$-decreases as compared to increases, with a ratio 
11:6. The probability that this preference is by chance, however, is 
rather high, of the order 0.09. This probability was calculated by 
means of a binomial probability function, assigning a 0.5 probability 
that the peak $\zeta (3)$-change will be positive and assuming 11 
positive chance ``hits'' out of 17 independent trials 
(since the $\zeta (3) >0$  
probability is 0.5  
the binomial distribution becomes symmetric with respect to chance 
hits, so identical results 
would be reached in case we assumed 6 negative  
chance ``hits'' out of 17 trials). 
For M-flaring regions, stronger $\zeta (3)$-decreases barely 
dominate, with a ratio 32:28. This time, however,  
the binomial probability that 
this result is random is $\approx 3 \times 10^{-14}$, so we can safely 
conclude that there is no clear preference of 
$\zeta (3)$-decreases over increases in 
case of M-flaring regions. For non-flaring regions, the 
peak $\zeta (3)$-changes are almost evenly divided between decreases and 
increases, with a ratio 146:147, that is again not random (the binomial
probability that this result is random, given the large sample size, 
is practically zero). Overall, it becomes clear that one 
cannot use the preflare (Figure \ref{pre_max}b) or the peak  
(Figure \ref{all_max}b) change in $\zeta (3)$ to 
assess the flaring productivity of an active region, meaning that 
flaring regions do {\it not} show distinguishable,  
sharp changes in their degree of intermittency.   

\begin{table}[t] 
\caption{Summary of means and standard deviations (in 
  parentheses) for the preflare and  
  peak values of $\Phi _{tot}$, the scale-free $D_0$, and 
  multiscale parameters $\alpha$ and $\zeta (3)$, 
  calculated in our sample of 17 X-flaring, 60 M-flaring, 
  and 293 non-flaring active regions. The preflare active-region values 
  correspond to 24 X-class and 87 M-class flares. The 
  discrepancy between numbers of flaring regions and flares is because 
  some flaring regions flare repeatedly over the observing interval.} 
\label{tab2} 
\begin{tabular}{lccccc} 
\hline 
 & \multicolumn{2}{c}{\large{Preflare values}} & \multicolumn{3}{c}{\large{Peak values}}\\ 
Active Regions & X-flaring & M-flaring & X-flaring & M-flaring & Non-flaring\\  
\hline 
Data points & 24 & 87 & 17 & 60 & 293\\ 
$\Phi _{tot}\;(\times 10^{22}\;Mx)$  & 4.57 (1.63) & 3.55 (1.72) & 4.74 (1.68) & 3.84 (1.56) & 2.58 (1.21)\\ 
 
$D_0$                & 1.44 (0.07) & 1.40 (0.08) & 1.51 (0.05) & 1.50 (0.05) & 1.44 (0.13)\\ 
$\alpha$            & 2.38 (0.29) & 2.40 (0.33) & 2.77 (0.36) & 2.73 (0.36) & 2.60 (0.42)\\ 
$\zeta (3)$-change  & 0.012 (0.06) & -0.001 (0.11) & -0.03 (0.20) & -0.04 (0.26) & -0.004 (0.29)\\ 
\hline 
\end{tabular} 
\end{table} 
Table \ref{tab2} quantifies Figures  
\ref{pre_max} and \ref{all_max}, providing means and standard 
deviations for each depicted distribution. It shows that only the 
peak values of the unsigned flux $\Phi _{tot}$ and the fractal 
dimension $D_0$ are somewhat different between flaring and non-flaring 
regions but the respective dispersions are such that there is  
considerable mixing of values between the three active-region 
populations. For the preflare values of $\Phi _{tot}$ and $D_0$ there 
is even more mixing and, in terms of $D_0$, X- and M-flaring regions 
are practically indistinguishable. When it comes to $\alpha$-values 
and changes in $\zeta (3)$, Table \ref{tab2} -- along with Figures 
\ref{pre_max} and \ref{all_max} -- shows that these multiscale 
parameters simply cannot be used to distinguish flaring from 
non-flaring regions, let alone predict large flares within a given 
time span.  
\subsection{Flare Forecasting Probabilities}
\label{S-flprob}
In Sections \ref{S-prepeak}, \ref{S-allpeak} we found that 
neither the preflare nor the  
peak values of our scale-free and multiscale parameters 
seem capable of distinguishing flaring from non-flaring regions. The 
unsigned flux $\Phi_{\mathrm{tot}}$ tends to score better than the 
scale-invariant $D_0$, the multiscale $\alpha$, and the change in 
$\zeta (3)$. To further quantify these results we calculate here  
the conditional probability of an active region being a flaring one if 
a given parameter inferred from any one of its magnetograms exceeds a 
preset threshold. A way to do this is by using Laplace's rule of 
succession and proceeding to a Bayesian inference of the predictive 
probability as follows: assume that $F$ magnetograms of flaring regions 
and $N$ magnetograms of non-flaring ones exhibit a value $R$ of a 
parameter that exceeds a threshold $R_{\mathrm{thres}}$. Then, the conditional 
probability $p$ that an active-region with $R > R_{\mathrm{thres}}$ will be a 
flaring one is given by (\opencite{Jaynes_03}, pp. 155--156)  
\beq 
p = {{F+1} \over {N+2}}\;\;\;\mathrm{with}\;\;\;\mathrm{uncertainty}\;\;\; 
\delta p = \sqrt{ {{p(1-p)} \over {N+3}} }\;\;. 
\label{cprob_eq} 
\eeq 
\bfig[t] 
\centerline{\includegraphics[width=1.\textwidth,clip=]{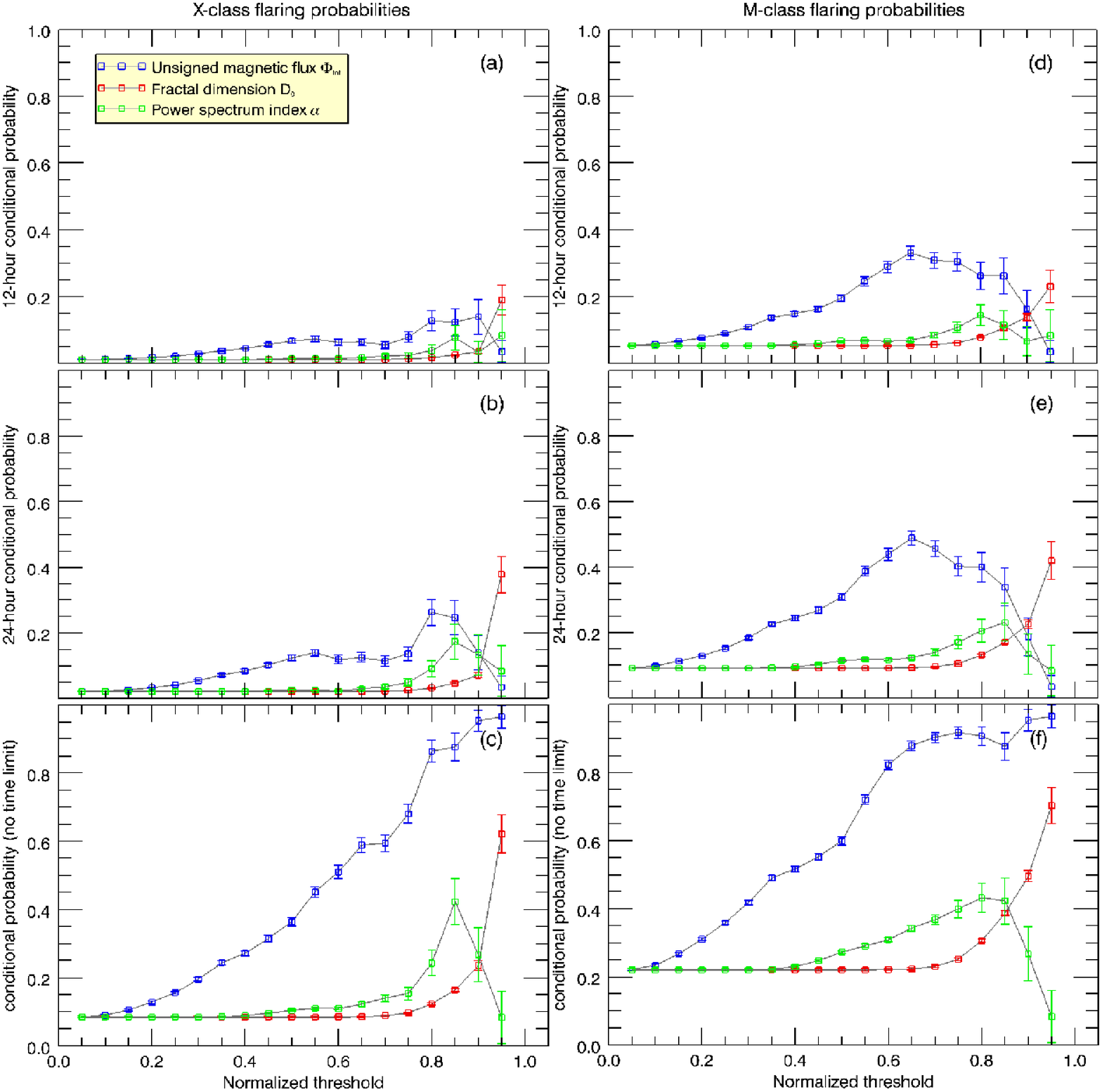}} 
\caption{Conditional probabilities of active regions to host a major 
  flare, either X-class (left column) or M-class (right column), with 
  respect to the normalized (against the maximum) threshold of a given 
  parameter, with $\Phi_{\mathrm{tot}}$ (blue squares), $D_0$ (red squares),  
  and $\alpha$ (green squares) examined. The upper (a, d) and middle 
  (b, e) rows provide the 12- and 24-hour conditional probabilities, 
  respectively. The lower row (c, f) provides the probability without 
  a time limit, that is, the probability of major flaring at a future 
  time when the active region is still visible in the disk.  
  The maximum values against which the thresholds were normalized are   
  $8.8 \times 10^{22}$ Mx, $1.65$, and $4.17$ for $\Phi_{\mathrm{tot}}$, 
  $D_0$, and $\alpha$, respectively.}  
\label{cprob} 
\efig 
This probability rule was also used by \inlinecite{Wheatland_05} to 
test another solar-flare prediction method. To compare directly  
between the conditional probabilities of the various 
parameters, we normalize the thresholds with respect to the maximum value 
of each parameter appearing in Figures \ref{pre_max} and \ref{all_max}. We 
examine only three of the four parameters 
- $\Phi_{\mathrm{tot}}$, $D_0$, and $\alpha$ -- because the 
changes of $\zeta (3)$ obviously  
show very similar patterns for flaring and non-flaring 
regions. Using the NOAA/GOES and {\it Yohkoh}/HXT flare catalogs (Section 
2.1), we consider a given active-region magnetogram as a preflare one 
if a major flare happened in the region within a given, preset 
timeframe from the time that the magnetogram was recorded. Figure 
\ref{cprob} shows conditional probabilities for two timeframes: 12 
hours (Figure \ref{cprob}a, d) and 24 hours (Figure \ref{cprob}b, e), 
while Figures \ref{cprob}c and f show conditional probabilities without 
any timeframe assigned: if an active region flared at any time  
after the magnetogram was taken, then this magnetogram is considered a 
preflare one. Notice that the flares under examination may have occurred
  when a given region has moved beyond the $30^o$ E-W meridional zone of
  analysis. This does not affect the analysis, however, as the preflare 
  magnetograms were acquired when the region was still within
  the analysis zone.  

All plots of Figure \ref{cprob} illustrate that the unsigned flux 
$\Phi _{tot}$, a conventional activity predictor, is generally more 
effective in predicting major flares than both the scale-free fractal 
dimension $D_0$ and the multiscale turbulent power-spectrum index 
$\alpha$. Differences between $\Phi _{tot}$ and ($D_0$, $\alpha$) are 
smaller (but also reflect small flare probabilities, of limited practical use)
in case of the most demanding prediction, the one with a 12-hour 
timeframe for flares $>$X1.0 (Figure \ref{cprob}a). In all other cases  
$\Phi _{tot}$ gives much higher (well beyond error bars)  
probabilities than $D_0$ and $\alpha$. The predictive 
ability of $D_0$ appears comparable with, or slightly higher than, that 
of $\Phi _{tot}$ only for normalized thresholds $R_{\mathrm{thres}} \ge 0.9$ 
for the 12- and 24-hour prediction timeframes. For the same timeframes,  
the predictive ability of $\Phi _{tot}$ drops for  
$R_{\mathrm{thres}} \gtrsim 0.8$ (Figures \ref{cprob}a, b) and  
$R_{\mathrm{thres}} \gtrsim 0.6$ (Figures \ref{cprob}d, e). This is because the 
upper $\Phi _{tot}$-ranges in these cases are occupied by non-flaring 
(within the preset timeframes) active regions -- this is one of the
limitations for using the unsigned flux as a flare predictor. 
The power-spectrum index 
$\alpha$ exhibits similar behavior, but for higher  
$R_{\mathrm{thres}} \gtrsim 0.8$, in all 
plots of Figure \ref{cprob}. Comparing the multiscale $\alpha$ with 
the scale-free $D_0$, we note that $\alpha$ works somewhat better, 
especially for larger timeframes. This is in line with 
\citeauthor{Abramenko_05}'s (\citeyear{Abramenko_05}) suggestion that 
$\alpha$ better reflects future flare productivity.   
However, recall that $\alpha$  
depends sensitively on the spatial resolution  
of the observing instrument, contrary to $D_0$ (Table \ref{tab1}). Hence  
the results of Figure \ref{cprob} concerning 
$\alpha$ should be viewed as holding exclusively for MDI low-resolution 
magnetograms.  
\section{Summary and Conclusion} 
\label{S-conc} 
This study investigates previous claims on the efficiency of 
fractal and multifractal techniques as reliable predictors of 
major solar flares and/or parameters reflecting the overall flare 
productivity of solar active regions before they actually flare.  
From the array of parameters implemented in the literature, we select 
three of the reported most promising ones: the 
fractal dimension, the multifractal intermittency index, and the  
scaling index of the turbulent power spectrum. Our 
objective is not to judge the methods {\it per se} but, rather, to test 
the notion of utilizing fractality and multifractality to gain 
predictive insight into major solar flares. 
 
Statistical analyses such as this one must guarantee that the assembled 
active-region sample is representative:  
the sample must contain numerous   
flaring {\it and} non-flaring regions. Comprehensive  
statistics often help avoid the interpretation of 
incidental signals as statistically significant behavior.  
Section~\ref{S-prepeak} (Figure \ref{pre_max}d)  
includes examples of results that 
might have been interpreted in a misleading way had the statistics 
of our active-region sample been insufficient.
 
We study 370 SOHO/MDI low-resolution ($1.98''$ per pixel)  
timeseries of active-region 
magnetograms, 293 of which correspond to active regions without 
major flares and 77 correspond to M- and X-class flaring 
regions. MDI line-of-sight fields are used for regions  
within $30^o$ of the central meridian in 
order to approximate the longitudinal-field component with the normal-field 
component and avoid any corrections or otherwise modifications 
of the original MDI data. 
We find that neither scale-free (fractal) nor multiscale 
(multifractal) techniques can be used to predict major flares,  
or for the {\it a priori} assessment of the flaring productivity of active 
regions. In particular, we find that their diagnostic capability 
is not better than that of the unsigned magnetic 
flux of active regions, a traditional, but unreliable, activity predictor.  
Since the fractal and multifractal measures tested here are less 
effective than the unsigned flux (Figure \ref{cprob}), they should not 
be used for flare prediction or for flaring productivity assessment. 
 
On the fundamental question of whether flaring active regions 
are more fractal, multifractal, or turbulent than other, non-flaring 
ones, the answer {\it per} our results has to be negative: flaring 
regions tend to exhibit relatively large peak values 
of scale-free and multiscale parameters but 
these values, or even higher ones sometimes,  
are also exhibited by non-flaring regions. For all statistical 
distributions, the means and standard deviations are such 
that the different populations of flaring and non-flaring regions 
overlap considerably (Table \ref{tab2}).      
 
At this point we emphasize our willingness to follow the guidelines 
of multiple previous studies in the inference of the above fractal and 
multifractal parameters. In particular, we followed 
\inlinecite{McAteer_etal05} when inferring the fractal dimension 
$D_0$, \inlinecite{Abramenko_05} when inferring the turbulent scaling 
index $\alpha$ (despite the fact that Abramenko worked exclusively 
on high-resolution MDI magnetograms), and a previous work of this 
author \cite{Georgoulis_05}, together with 
\inlinecite{Abramenko_etal03}, when inferring the intermittency index  
$\zeta (q)$. As a result, the findings of  
both \inlinecite{McAteer_etal05} and  
\inlinecite{Abramenko_05} were qualitatively reproduced in this 
analysis, while we 
showed that the distinct $\zeta (3)$-behavior reported by 
\inlinecite{Georgoulis_05} was just one incidental case and not part 
of a systematic trend.  
 
In addition, this work (Section \ref{S-spres}) exposes a dependence of
multiscale parameters $\zeta (q)$ and $\alpha$ on the spatial resolution of
the studied magnetograms. In contrast, the scale-free $D_0$ appears fairly
insensitive to varying spatial resolution. Therefore, results and comparisons
for $\zeta (3)$ and $\alpha$ in Section \ref{S-comp} are valid only for MDI
low-resolution data and should not be generalized to data sets of other
instruments. Possible susceptibility of the $D_0$-value should also be 
studied with respect to the threshold it requires, unlike $\zeta (3)$ and
$\alpha$. This investigation has not been carried out here. In previous works,
however, \inlinecite{Meunier_99} reported a decreasing trend of $D_0$ with
increasing threshold, while \inlinecite{Janssen_etal03} reported a slighter  
decrease, or a near insensitivity, of $D_0$ for increasing thresholds,
in case these thresholds are sufficiently above noise
levels or the magnetic field data have been treated for noise, respectively.

It is useful to mention here a very recent result by  
\inlinecite{Abramenko_Yurchyshyn10} that the turbulent power-spectrum 
index $\alpha$, either alone or coupled with the integral of the 
power-spectrum for all wavenumbers, correlates better than 
$\Phi_{\mathrm{tot}}$ with the flaring index in a large sample of 217 active 
regions recorded in high-resolution MDI magnetograms. While 
correlating some parameter with the flaring index is not identical to 
inferring the predictive capability of this parameter, these results 
appear in likely contrast with the results presented here.  
Further investigation is clearly needed, therefore. 
Nonetheless, some convergence of views appears 
in that multiscale parameters may not be 
ideal tools for solar flare prediction (Abramenko, 2010, private 
communication).    
 
Perhaps more instructive than pointing out the inability of scale-free 
and multiscale techniques to assess {\it a priori} 
the flaring record of active 
regions is to explain {\it why} this is the case. In this author's 
view, there are at least two distinct reasons that justify our findings:  
 
First, fractality and multifractality are extremely 
widespread in the solar atmosphere, eruptive and quiescent alike. This 
may well be due to the turbulence dominating the magnetic-flux 
generation and emergence process (see Introduction). For example, recall 
the fractality of white-light granules  
\cite{Roudier_Muller87,Hirzberger_etal97},  
the fractality and  
multifractality of active regions and the 
quiet-Sun magnetic field  
\cite{Schrijver_etal92,Cadavid_etal94,Meunier_99,Janssen_etal03},  
the fractality of flares and sub-flares in the EUV  
(\opencite{Aschwanden_Parnell02}; 
\opencite{Aschwanden_Aschwanden08a}; \citeyear{Aschwanden_Aschwanden08b}),  
the fractality of the quiet network in the EUV 
\cite{Gallagher_etal98},  
that of Ellerman bombs in off-band H$\alpha$ 
\cite{Georgoulis_etal02}, and others. The fractal dimension in most, 
if not all, of these works varies between 1.4 and 1.8, 
practically indistinguishable from the fractal dimension of active 
regions found here. As a result, it appears unlikely that these 
same methods may reflect particular characteristics 
of active regions, let alone flare productivity.  
 
Second, there is a lack of correlations between the 
fractal dimension in the photosphere and that of the overlying 
chromosphere and corona, where major flares occur. 
\inlinecite{Dimitropoulou_etal09}  
assumed nonlinear force-free magnetic fields extending 
above the photosphere and calculated volumes of enhanced electric 
currents and steep magnetic gradients from these extrapolated 
fields. They found no correlation between the three-dimensional 
fractal dimension of these volumes and that of the two-dimensional 
photospheric boundary. In other words, all photospheric ``memory'', in terms
of fractality and multifractality, is erased above the photosphere due to the
fact that these unstable volumes become nearly space-filling slightly above
this boundary. Attempting to assess the fractality of 
layers higher than the photosphere -- where flares occur -- by using the 
photospheric fractality as a proxy will not yield meaningful results,  
similarly to the lack of correlation between 
photospheric electric currents and the coronal X-ray brightness 
\cite{Metcalf_etal94}.  
 
In addition, it is possible that both flaring and non-flaring regions  
share a similar degree of 
self-organization in the distribution of their magnetic free energy, 
as reported by \inlinecite{Vlahos_Georgoulis04}. 
Flaring regions have an ``opportunity'' to show their  
self-organization via flaring, with flares inheriting  
the statistics of their host active regions,  
while non-flaring regions retain this property without 
demonstrating it. In this sense {\it i)} fractality alone  
cannot be responsible for flaring, and  
{\it ii)} fractality, as a global characteristic of the active-region 
atmosphere, cannot be used to determine {\it a priori} which active regions 
will flare.   
 
There are, of course, sophisticated multiscale techniques not treated 
in this work, such as wavelet methods used to extract the magnetic-energy 
spectrum in active regions \cite{Hewett_etal08} or to distinguish 
active regions from quiet Sun for further treatment \cite{Conlon_etal10}, 
or the flatness function and its 
intermittency index \cite{Abramenko_etal08}. 
While we cannot comment on methods that we have not tested, per our
conclusions it would seem rather surprising if a scale-free or 
multiscale technique delivered a notable improvement in our 
forecasting ability, as this would apparently contradict what 
scale-free and multiscale behavior caused by self-organization  
is meant to imply: spontaneity in 
the system's dynamical response to external forcing, both in timing 
and in amplitude, and hence a lack of certainty in predicting this 
response.  
 
Let us finally mention that alternative flare prediction 
approaches have been developed in recent years. Rather than fractality, 
multifractality, or intermittency and turbulence, these methods rely 
on parameters stemming from morphological and topological  
characteristics of active regions, such as those  
of the photospheric magnetic-polarity inversion lines  
or photospheric properties in general 
(\opencite{Falconer_etal06}; \opencite{Schrijver_07};  
\opencite{Georgoulis_Rust07}; \opencite{Leka_Barnes07};  
\opencite{Mason_Hoeksema10}), or those of the  
subsurface kinetic helicity prior to active-region emergence 
\cite{Reinard_etal10}, among others. It remains to be seen 
whether these parameters can lead to advances in the forecasting of 
major solar eruptions or whether forecasting will remain inherently 
probabilistic which, per our results, seems entirely possible. In any 
case, fractal and multifractal methods -- perhaps not extremely useful 
as eruption predictors -- will always be   
excellent tools for a fundamental understanding 
of the origins and nature of solar magnetism.  
\begin{acks} 
This work is based on a talk given by the author  
during the Fourth Solar Image Processing (SIP) Workshop in   
Baltimore, MD, USA, 26-30 October 2008. Thanks are due to the 
organizers for an interesting and productive meeting.  
During the author's tenure at the Johns Hopkins 
University Applied Physics Laboratory (JHU/APL)  
in Laurel, MD, USA, this work 
received partial support from NASA's LWS TR\&T Grant   
NNG05GM47G and Guest Investigator Grant NNX08AJ10G. The author  
gratefully acknowledges the Institute of Space Applications and Remote 
Sensing (ISARS) of the National Observatory of Athens for the 
availability of their computing cluster facility for massive runs 
related to this work. SOHO is a project of international cooperation 
between ESA and NASA. {\it Hinode} is a Japanese mission developed and 
launched by ISAS/JAXA, with NAOJ as domestic partner and NASA and STFC 
(UK) as international partners. It is operated by these agencies in 
co-operation with ESA and NSC (Norway). Finally, the author thanks 
the two anonymous referees for contributing to the clarity, accuracy, and
focus of this work.   
\end{acks} 
%
\bibliographystyle{spr-mp-sola} 
 
\bibliography{SOLA_ms_references}   
 
%
\end{article}  
\end{document}